\documentclass[11pt,superscriptaddress]{revtex4}
\usepackage{amsmath,amssymb,graphicx}
\usepackage[usenames]{color}
\usepackage{multirow}
\usepackage{array}
\usepackage{longtable}
\usepackage{graphicx}
\usepackage{url}
\usepackage{graphicx,epstopdf}
 \fontsize{15}{15}
\newcommand\comment[1]{}

\newcommand\red{\color{red}}
\newcommand\blu {\color{blue}}
\newcommand\beq {\begin{equation}}
\newcommand\eeq {\end{equation}}

\begin{document}

\centerline{\large\bf Inertio-elastic instability in
Taylor-Couette flow of a model wormlike micellar system } \vskip 5
mm \centerline{Hadi Mohammadigoushki$^{1}$ and Susan J. Muller
$^{2*}$} \vskip 5 mm

\centerline{$^1$Department of Chemical and Biomedical Engineering}
\centerline{Florida A\&M University - Florida State University,
College of Engineering, Florida 32310, United
States} 
\vskip 5 mm \centerline{$^2$Department of Chemical and
Biomolecular Engineering} \centerline{University of California,
Berkeley, California 94720, United States} \vspace*{0.1 in} {

\vskip 5 mm

\centerline{\large\bf Abstract}\vskip 5 mm

In this work, we use flow visualization and rheometry techniques
to study the dynamics and evolution of secondary flows in a model
wormlike micellar solution sheared between concentric cylinders,
i.e., in a Taylor-Couette (TC) cell. The wormlike micellar
solution studied in this work contains cetyltrimethylammonium
bromide (CTAB) and sodium salicylate (NaSal). This system can be
shear banding and highly elastic, non-shear banding and moderately
elastic, or nearly Newtonian as the temperature is varied over a
narrow range. The effect of elasticity on transitions and
instabilities is probed by changing the temperature over a wide
range of elasticity (El $\ll$ 1, El $\approx$ 1, and El $\gg$ 1).
Elasticity is defined as the ratio of the Weissenberg number to
the Reynolds number. For shear banding wormlike micelle {{\red
solutions}} where El $\gg$ 1, a primary transition from the base
Couette flow to stationary vortices that are evenly spaced in the
axial direction of the shear cell and are characterized by an
asymptotic wave-length is observed. The dimensionless wave-length
at the onset of this shear banding transition for CTAB/NaSal
system turns out to be much larger than those reported for other
shear banding wormlike micelle systems. For the same fluid at a
temperature where it shear-thins but does not display shear
banding, El $\approx$ 1, and for slow ramp speeds, the primary
transition is to distinct structures that are not stationary but
rather travel in the axial direction. At low elasticity (El $\ll$
1), where the fluid behaves as a nearly Newtonian fluid, several
transitions from purely azimuthal Couette flow to modified Taylor
vortex flows and finally chaotic regimes are documented. The
behavior in the shear-banding and non-shear-banding regimes are
discussed and compared with results in related systems. The
possibility of hysteresis in the flow transitions as well as the
effects of co-rotation and counter-rotation of the cylinders on
transitions and instabilities are also examined for a wide range
of elasticity.

\newpage
\section{Introduction}

The hydrodynamic stability of Taylor-Couette flow, or flow between
two concentric cylinders, has received considerable attention
since the seminal work of G. I. Taylor~\cite{Ta23}. The
Taylor-Couette geometry is a versatile platform to study
instabilities in flowing fluids and is characterized by the
cylinder height $h$, inner cylinder radius $R_{i}$, and the outer
cylinder radius $R_{o}$. In non-dimensional terms the geometry is
parameterized in terms of an aspect ratio $\Gamma=h/(R_{o}-R_{i})$
and either a radius ratio $R_{i}/R_{o}$ or a curvature ratio
$\varepsilon = (R_{o}-R_{i})/R_{i}$. G. I. Taylor presented a
{{\red groundbreaking}} paper on the inertial instability of
Newtonian fluids sheared in the concentric cylinders geometry and
identified a critical threshold for the onset of the primary
{{\red inertial}} instability; in the case of rotation of the
inner cylinder only, this is now typically written as $Ta_{c}
=\varepsilon Re_{i}^{2}$, where $Re_{i}$ is a Reynolds number
based on the inner cylinder. $Re_{i}$ is defined as $ Re_{i} =
\rho R_{i} \Omega_{i}(R_{o} - R_{i}) /\eta$, where, $\rho$,
$\Omega_{i}$, and $\eta$ are the density of the fluid, angular
velocity of the inner cylinder, and viscosity of the Newtonian
fluid, respectively. This instability is thus related to a
coupling between inertial forces and streamline
curvature~\cite{Ta23}. Beyond this critical threshold the purely
azimuthal base flow is replaced by an axisymmetric, time
independent toroidal vortex flow, now referred to as Taylor Vortex
Flow (TVF). Several other transitions follow as $Re_{i}$ (or $Ta$)
is further increased, and eventually a Newtonian fluid {{\red
exhibits}} a series of turbulent states. In a subsequent {{\red
important}} study, Andereck and co-workers examined the effect of
co-rotation and counter rotation of the cylinders on the Newtonian
transitions and reported a wide variety of flow states and
transitions~\cite{An86}. Many excellent reviews on {{\red
Newtonian Taylor-Couette instabilities}} exist and readers are
encouraged to consult those~\cite{Sw81,Ta94,Fa14}.\newline

There is also a rich literature on transitions and instabilities
of viscoelastic polymeric fluids in the Taylor-Couette
geometry~\cite{La92,La94,La90,Mu89,Ba99,Wh00,Cr02,Du11,Du09b,Gr96,Gr98a,Gr98b,Gr97,St98,Gr93,Du13,Mu08}.
Researchers have shown that addition of a small amount of polymer
to Newtonian fluids dramatically alters the instabilities and
transitions reported for Newtonian fluids in the Taylor-Couette
{{\red geometry}}~\cite{Gr96,Gr98a}. Larson et al.~\cite{La90}
discovered a purely elastic instability in polymer solutions at
vanishing Reynolds numbers. They proposed a criterion for the
onset of elastic instability for an Oldroyd-B fluid based on a
linear stability analysis as follows: \beq K_{c} \sim
\varepsilon^{1/2} f(\eta_{p}/\eta_{s}) Wi, \label{eq1}\eeq where
the Weissenberg number $Wi$ is defined as the polymer relaxation
time multiplied by the shear rate. $\eta_{p}$ and $\eta_{s}$ are
the polymeric contribution to the viscosity and the solvent
viscosity, respectively. This purely elastic instability is thus
related to a coupling between elastic forces and streamline
curvature. This work inspired many other studies that investigated
different aspects of elastic instabilities of polymer solutions in
the Taylor-Couette
cell~\cite{Ba99,Cr02,Gr96,Gr98b,Gr97,St98,Wh00}. In addition,
there have been many efforts to study the inertio-elastic
instability in polymer solutions where both Reynolds and
Weissenberg numbers are important. One naturally expects the
critical threshold for onset of inertio-elastic instability ($S$)
in polymeric fluids to depend on several parameters; at a minimum
one would anticipate the following: \beq S \sim f(\varepsilon,
\eta_{p}/\eta_{s}, Wi, Re). \label{eq2}\eeq The competition
between elastic and inertial forces can be cast in a dimensionless
elasticity number defined as $El = Wi/Re$. In the range of very
low elasticity (i.e. $El \ll 1$), researchers have primarily
recovered transitions similar to Newtonian fluids, with the
critical thresholds shifted slightly due to the presence of a
small amount of elasticity~\cite{Cr02,Du11}. For example, Dutcher
and Muller studied a dilute solution of polyethylene oxide (PEO)
in glycerol/water and reported the same transition sequence as for
Newtonian fluids, namely Couette flow (CF) to axisymmetric time
independent vortex flow (TVF) to wavy vortex flow (WVF) for a
range of $10^{-4} < El < 0.02$, and $0.3 < \eta_{p}/\eta_{s} <
0.93$ [13]. {{\red Similar results have been reported by
Crumeyrolle and co-workers~\cite{Cr02} and Groisman and
Steinberg~\cite{Gr93,Gr98b}, although the latter authors report a
modified transition sequence under certain conditions of $El$ and
$\eta_{p}/\eta_{s}$.}}\newline

In the range of moderate elasticity ($El \approx O(1)$), the
transitions {{\red observed for Newtonian fluids}} are modified by
elasticity and replaced by new transitions~\cite{Du09b,Cr02}.
{{\red As an example, f}}or PEO/water solutions with PEO
concentrations above 500 ppm, Crumeyrolle et al. reported a
primary transition from Couette flow to {{\red a new}} standing
wave (SW) {{\red state}} in the range of $0.07<El<0.5$, and
$5.32<\eta_{p}/\eta_{s}<12.4$~\cite{Cr02}. More recently, Dutcher
and Muller reported {{\red a different series of (non-Newtonian)
transitions for a slightly shear-thinning PEO solution for
$El\approx$ 0.1-0.2 and found the transition sequence was
hysteretic~\cite{Du13}.}}  Groisman and Steinberg also studied
instabilities in Taylor-Couette flow of a solution of
polyacrylamide (PAAm) in a viscous sugar syrup (aqueous solutions
of saccharose of different concentrations) systematically {{\red
and observed transitions at low $El$ that were dependent on $El$
and, for $El>0.3$ were hysteretic}}~\cite{Gr96,Gr98b,Gr97,St98}.
Further detailed information on inertio-elastic instabilities in
polymer solutions can be found in recent
reviews~\cite{St98,Mu08}.\newline

Most of the above experiments on viscoelastic polymer solutions
have focused on inner cylinder rotation while the outer cylinder
was held fixed. However, as in Newtonian fluids, the simultaneous
rotation of both cylinders has a dramatic effect on transitions
reported in polymeric fluids~\cite{Du11,Du13}. Dutcher and Muller
studied the effect of co- and counter rotation of cylinders over a
modest range of elasticity for polymer solutions. For a dilute
solution of PEO in water/glycerol and in the range of low
elasticity $El \approx 0.023$, increasing co-rotation of the
cylinders shifted the critical Reynolds number for the onset of
instabilities to higher values. Moreover, Dutcher and Muller
showed that for the range of moderate elasticity ($El \approx$
0.1-0.2), when cylinders rotate in counter fashion the {{\red
primary transition from CF is to disordered rotating standing
waves (DRSW) rather than to stationary vortices}} at high $Re_{o}$
~\cite{Du11,Du13}.\newline

Recently, surfactant based viscoelastic fluids (i.e. wormlike
micellar fluids) have been widely used as model viscoelastic
solutions for rheological studies~\cite{Le09}. These fluids are
made by dissolving surfactants and salts in water. In contrast to
polymer solutions that exhibit a spectrum of relaxation times,
some wormlike micellar solutions follow a single mode Maxwell
model in their linear viscoelastic behavior~\cite{Re91}. In the
non-linear regime, they {{\red sometimes}} exhibit shear banding
which is characterized by the formation of a plateau in the flow
curve within a range of shear rates~\cite{Le09,He09,Mo16a,He09b}.
In recent years, wormlike micelles have been used to study
viscoelastic instabilities in different geometries such as
Taylor-Couette cells, cross slots, microchannels and also in flow
past an obstacle~\cite{Le08,Du12,Mo16a,Ha12,Ng10,Mo16b}. In
Taylor-Couette geometries, most of these studies are focused on
the purely elastic instability and rotation of the inner cylinder
only~\cite{Fa12a,Fa12b,Le06,Le08}. Taylor-Couette studies by
Fardin and co-workers have shown, for a broad range of
shear-banding wormlike micelle systems in the high elasticity
limit, {{\red a primary transition to}} Taylor-like vortices in
the high shear band that cause{{\red s}} the interface between the
low and high shear bands to undulate. Near the critical
conditions, these vortices oscillate in the axial direction and
are characterized by a wave-length that scales with the thickness
of the high shear band~\cite{Fa12a,Le06,Le08,Fa12b,Mo16a}.
\newline

To the best of our knowledge, there is only one work that has
reported an inertio-elastic instability for wormlike micelle
{{\red solutions}} in the Taylor-Couette geometry, and there is as
yet little data on flow transitions and instabilities over a wide
range of elasticity in {{\red these solutions}}~\cite{Pe14}. Perge
et al. studied instabilities and transitions in a non-shear
banding wormlike micelle system based on CTAB/NaNO$_{3}$ (0.1 M,
0.3 M) at moderate elasticity ($El\approx 1$) and rotating only
the inner cylinder{{\red~\cite{Pe14}}}. Following a {{\red
step-wise}} increment in $Re_{i}$ they reported transitions
similar to transitions reported for polymer solutions based on
PEO/glycerol/water studied by Dutcher and Muller in the range of
moderate elasticity $El\approx$ 0.1-0.2~\cite{Du13}. Perge et al.
also showed that the primary transition is supercritical, and they
observed no significant hysteresis for any of the transitions in
their system. In contrast, Dutcher and Muller reported hysteresis
in the critical conditions and flow states for viscoelastic
solutions based on PEO/glycerol/water for $El\approx$
0.1-0.2~\cite{Du13}.\newline

The main objectives of this work are as follows. First, we wish to
examine systematically the inertio-elastic instability in a
wormlike micellar fluid by varying the elasticity number over a
wide range and examining any similarities or differences with
results reported for polymer solutions. In contrast to polymeric
solutions, wormlike micellar fluids are not susceptible to
mechanical degradation and are much easier to prepare. Wormlike
micelles can break and reform and this potentially leads to some
differences with transitions reported for polymer solutions. This
makes wormlike micelles perfect candidates to study hysteresis in
Taylor-Couette flow. However, shear banding wormlike micelle
{{\red solutions}} studied in the past {{\red we}}re highly
elastic and Taylor-Couette experiments always showed dominantly
bulk, purely elastic instabilities. To vary the elasticity in
wormlike micelle {{\red solutions}} our approach is to change the
temperature of the solution; in this regard, our approach is
similar to that of Groisman and Steinberg~\cite{St98} for polymer
solutions. We note, however, that as we vary the temperature, we
also vary the rheological behavior of the solutions, from
shear-banding (at $El \gg 1$) to shear-thinning with no
shear-banding to nearly Newtonian behavior ($El \ll 1$). Most of
the wormlike micellar solutions studied in the past are
thermo-thinning~\cite{Ch13}. Increasing the temperature reduces
the persistence length of the wormlike micelles and consequently
the bulk viscosity and elasticity. However, it is quite
challenging to find a wormlike micellar solution that allows us to
{{\red access}} a wide range of rheological behaviors within the operating
temperature range of our Taylor-Couette cell (20$^{\circ}$C $<$ T
$<$ 40$^{\circ}$C). It turns out that the wormlike micellar system
based on CTAB/NaSal (0.075M, R =[NaSal]/[CTAB]= 0.32) studied by
Dubash et al.~\cite{Du12} allows us to study instabilities in a
wide range of elasticity ( $O(10^{-3})< El < O(100)$ ) within a
narrow range of temperature (22$^{\circ}$C $<$ T $<$
35$^{\circ}$C). This enabled us to observe several interesting
flow transitions in wormlike micelle {{\red solutions}} that have
not been reported before, and to compare critical conditions to
those in polymer solutions. Second, we note that we can introduce
co- or counter-rotation of the cylinders. This broadens the range
of flow transitions we can examine, and, in the case of counter
rotation, allows us to change the characteristic length scale of
the flow by introducing a nodal surface between the cylinders.
This may potentially shift the thresholds of instabilities in
flows of wormlike micelle {{\red solutions}}. To the best of our
knowledge, the effects of co and counter rotation have not been
studied for wormlike micellar systems over a wide range of
elasticity.

\section{Experiments}

\subsection{Materials and Methods}

The wormlike micell{{\red e}} solution studied in this work is an
aqueous solution of cetyltrimethylammonium bromide (CTAB) and
sodium salicylate (NaSal) (both supplied by Fischer Scientific).
{{\red S}}olutions of 0.075 M CTAB and NaSal in de-ionized water
with a ratio of [NaSal]/[CTAB] = 0.32 {{\red were prepared and
mixed }}with a magnetic stir bar. {{\red Solutions were}} left at
room temperature for a few days to {{\red allow}} their
equilibrium structure {{\red to form}} {{\red prior to flow}}
experiments. For experiments in the Taylor-Couette cell a small
amount of mica flakes (0.001 wt\%{{\red , Mearlin superfine, Mearl
Corporation}}) {{\red were added}} to {{\red facilitate
visualization of}} the flow field. This small amount does not
affect the rheology of the solution. Mica flakes are anisotropic
particles that orient in the direction of flow, producing
variations in the intensity of light reflected by ambient lighting
and allowing us to monitor the formation and movement of vortex
boundaries. Abcha and co-workers~\cite{Ab08} have demonstrated
that in the Taylor-Couette system the intensity of reflected light
is directly related to the magnitude of the radial velocity
component.
\begin{figure}[h!]
  \centering
    \includegraphics [width=0.63\textwidth]{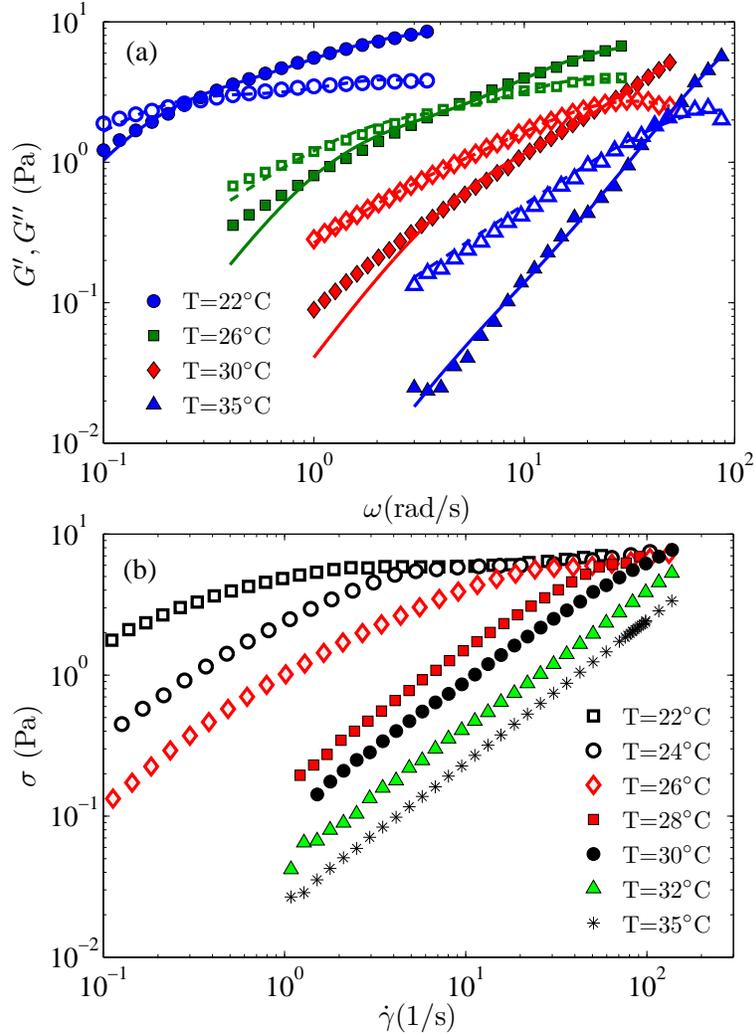}
   \caption{\small {{\red Rheology of surfactant micelle solutions: (a)}} Storage (filled symbols) and loss moduli (open symbols) versus angular frequency.
   Curves show predictions of a two mode Maxwell model. (b) Shear stress versus shear rate for the range of temperature studied in this work {{\red measured in a concentric cylinders geometry}}.
   {{\red Flow curves were obtained in strain controlled mode and for solutions the interval between each data point was set to 400 seconds to make sure the steady state is
   achieved.}}}
   \label{fig1}
  \end{figure}
{{\red A Malvern Gemini stress-controlled rheometer was used t}}o
characterize the rheology of the wormlike micell{{\red e
solution}} at different temperatures, {{\red in}} both linear and
non-linear viscoelastic shear experiments. The linear and
non-linear tests are carried out in a Couette co-axial cylinders
geometry with $R_{i}$ = 1.25 cm and $R_{o}$ = 1.375 cm. Flow
measurements are carried out in a custom built, computer
controlled Taylor-Couette cell that allows us to independently
rotate both cylinders. The radii of the inner and the outer
cylinders are 6.946 cm and 7.615 cm respectively, which provides a
radius ratio of 0.91. The height of the Taylor-Couette cell is
40.6 cm corresponding to an aspect ratio of 60.7 that renders end
effects negligible. The Taylor-Couette cell used in this study has
been extensively used in the past for studies of flow transitions
and instabilities~\cite{Ba99,Du09b,Wh00}. Images of the variations
in the light intensity in the $\theta-z$ plane of the
Taylor-Couette cell are captured using a CCD camera (model COHU
4910{{\red, COHU, Inc.}}) under ambient light exposure. Details of
the Taylor-Couette cell, image capture, and image processing may
be found in~\cite{Dut09}.
\section{Results and Discussion}
\subsection{Fluid Characterization}
The system of wormlike micelles based on CTAB/NaSal used in this
work has been studied by other researchers in microfluidic
studies~\cite{Du12}. This solution is very sensitive to small
variations in temperature. Small amplitude oscillatory shear
experiments were performed to extract the longest relaxation time
$\tau_{M}$ of the fluid. Fig.~\ref{fig1}(a) shows typical linear
viscoelastic results along with the prediction of a two mode
Maxwell model (eq.~(\ref{eq3}) and~(\ref{eq4})) at different
temperatures. We note that a single mode Maxwell model does not
fit well {{\red the measured fluid response}}. This is consistent
with the findings of Dubash et al. for the linear viscoelastic
results of solutions based on CTAB/NaSal in the concentration
range studied in this work~\cite{Du12}. \beq
G'(\omega)=\sum_{i=1}^{2}\frac{{\tau_{i}}^{2}\omega^{2}}{1+{\tau_{i}}^{2}\omega^{2}}~G_{oi}
\label{eq3}\eeq

\beq
G''(\omega)=\sum_{i=1}^{2}\frac{{\tau_{i}}\omega}{1+{\tau_{i}}^{2}\omega^{2}}~G_{oi}
\label{eq4}\eeq Similarly, it has been shown that {{\red other}}
surfactant-based viscoelastic solutions depart from ideal, single
mode Maxwell behavior as the concentration decreases from well
inside the semi-dilute regime towards the dilute
regime~\cite{Be93}. Steady shear experiments were also performed
to measure the shear stress $\sigma$ as a function of the shear
rate at different temperatures (cf. Fig.~\ref{fig1}(b)). At the
lowest temperature tested in this work (T = 22$^{\circ}$C), the
shear banding plateau is evident, however increasing the
temperature causes the wormlike micelle {{\red solution}} to
become {{\red a}} non-shear banding viscoelastic fluid over an
intermediate range of temperature (26$^{\circ}$C - 32$^{\circ}$C)
and eventually to behave as a nearly Newtonian fluid
(non-shear-thinning, and only very weakly elastic) at the maximum
temperature (T = 35$^{\circ}$C). {{\red At T = 22$^{\circ}$C, the
lower and the upper limits of the shear banding plateau are
$\dot\gamma_{l} = $ 3.25 $s^{-1}$ and $\dot\gamma_{h} = $ 14.2
$s^{-1}$, respectively. At T = 24$^{\circ}$C, these values are
$\dot\gamma_{l} = $ 9 $s^{-1}$ and $\dot\gamma_{h} = $ 30.75
$s^{-1}$. This is best shown when the data in Fig.~\ref{fig1} (b)
are replotted on semi-log coordinates (cf. Fig. S1
in~\cite{Sup}).}}
\begin{table}[t]
  \centering
      \begin{tabular}{l*{5}{c}r}
      \hline
       T ($^{\circ}$C)& $\tau_{M}$~(s)& $\eta_{0}$~(Pa.s) & $\eta_{p}$/$\eta_{s}$ & $El$\\
      \hline\hline
       22   & $4.52$ & $17.4$  ~& ${{\red 850-1340}}$ ~& ${{\red 15.5-18.6}}$ \\

       24   & $2.55$ & $2.3$ ~& ${{\red 400-600}}$ &~ ${{\red 3.3-4.2}}$ \\

       26   & $0.88$ & $1.2$ ~& ${{\red 270-380}}$ &~ ${{\red 0.72-0.97}}$ \\

       28   & $0.54$ & $0.16$ ~& ${{\red 130}}$ &~ $0.2$ \\

       30  & $0.165$ & $0.08$ ~& ${{\red 70}}$ & ~${{\red 0.034}}$ \\

       32   & $0.11$ & $0.04$ ~& $39$ &~$0.01$ \\

       35   & $0.04$ & $0.025$  ~& $23$ &~ $0.003$ \\

      \end{tabular}
  \caption{\small List of surfactant solution properties for the range of temperatures tested in this work.}
  \label{table}
\end{table}\newline

The fluid properties, including the longest Maxwell relaxation
time $\tau_{M}$, the zero shear rate viscosity $\eta_{0}$, the
power law index n, viscosity ratio $\eta_{p}/\eta_{s}$ and the
Elasticity number $El=Wi/Re_{i}$, as a function of temperature T
are summarized in Table~\ref{table}. {{\red $\eta_{s}$ represents
the solvent (water) viscosity ($10^{-3}$ Pa.s).
$\eta_{p}(\dot{\gamma})$ is defined as the contribution of the
wormlike micelles to the surfactant solution viscosity
($\eta_{p}(\dot{\gamma})$ = $\eta(\dot{\gamma}) - \eta_{s}$).
$\eta(\dot{\gamma})$ is the total wormlike micelles solution
viscosity measured in the rheometer.
 Here we define the
Weissenberg number $Wi$ as $Wi=\tau_{M}\dot{\gamma}_{i}$ where
$\dot{\gamma}_{i}$ is the imposed shear rate, calculated as
$\dot{\gamma}_{i} = |R_{o}\Omega_{o} - R_{i} \Omega_{i}|/d$}}. The
Reynolds number based on the inner cylinder is defined as $Re_{i}
= \rho~\Omega_{i} R_{i} d/\eta(\dot{\gamma}_{i})$, where $d$ is
the gap between the cylinders, $d = R_{o}- R_{i}$. {{\red The full
results of our two-mode Maxwell model fit are included in Table SI
of supplementary material. We note that for all temperatures below
28$^{\circ}$C, the viscosity is shear rate dependent, and hence
$El$ varies with imposed shear rate; the range of $El$ accessed in
the present work is indicated in Table I.}}\newline

As noted above, the characteristic shear rate is defined as the
imposed shear rate, consistent with earlier studies and with the
rheological data. However, most of the solutions studied in the
present work are shear thinning or shear banding and this
definition does not account for the effect of shear thinning on
the base velocity field. Although the imposed shear rate provides
a good approximation for the flow of the shear banding fluid at
the beginning of the shear banding plateau (cf. Fig.~S3 in
~\cite{Sup}), for higher shear rates that are within the shear
banding regime, the imposed shear rate underestimates the local
shear rate in the high shear band. On the other hand, the shear
rate profile based on the shear thinning power-law fluid often
overestimates the shear rate in the high shear band (cf. Fig. S4
in ~\cite{Sup}). Thus, while the imposed shear rate is not
necessarily equal to the local shear rate at the inner cylinder,
we use it to be consistent with the literature results and the
rheological measurements.

\subsection{Flow Transitions and Instabilities}
In this section, we present results that illustrate the flow
transitions and instabilities for our wormlike micellar fluid in
the Taylor-Couette cell as the temperature, and hence $El$ and
rheological properties, are varied. {{\red The inner cylinder
rotation is slowly ramped at a rate of 0.001 rad/$s^{2}$ in all experiments, so that
flows at increasing $Re_{i}$ and $Wi$ are accessed for $Re_{o} =
0$ (that is, the outer cylinder is held stationary). We performed many experiments to establish
that this ramp rate is sufficiently slow that neither the critical conditions nor the flow state observed
were changed by using a slower ramp rate.}} Note that since the viscosity is shear rate dependent
(at all except for the highest temperatures), the Elasticity number varies modestly
during the ramp. {{\red Images of the}} $\theta-z$ plane of the TC
geometry {{\red are captured}} using a CCD camera. A single line
of pixels is selected in the axial direction of the TC cell for
each shear rate and collected over time (as $Re_{i}$ is slowly
increased) to form space-time (or space-$Re_{i}$) plots. The
space-time plot is a useful way to determine the $Re_{i}$
corresponding to the onset of flow transitions and instabilities,
and has been used extensively in the
literature~\cite{Cr02,Du09b,Gr96}. In the following, we present
such results for the same wormlike micellar solution as it
displays shear banding (and high $El$), shear-thinning
viscoelastic behavior (at moderate $El$), and nearly Newtonian
behavior (at $El\ll1$) in steady shear rheology.
\begin{figure}[h!]
  \centering
    \includegraphics [width=0.9\textwidth]{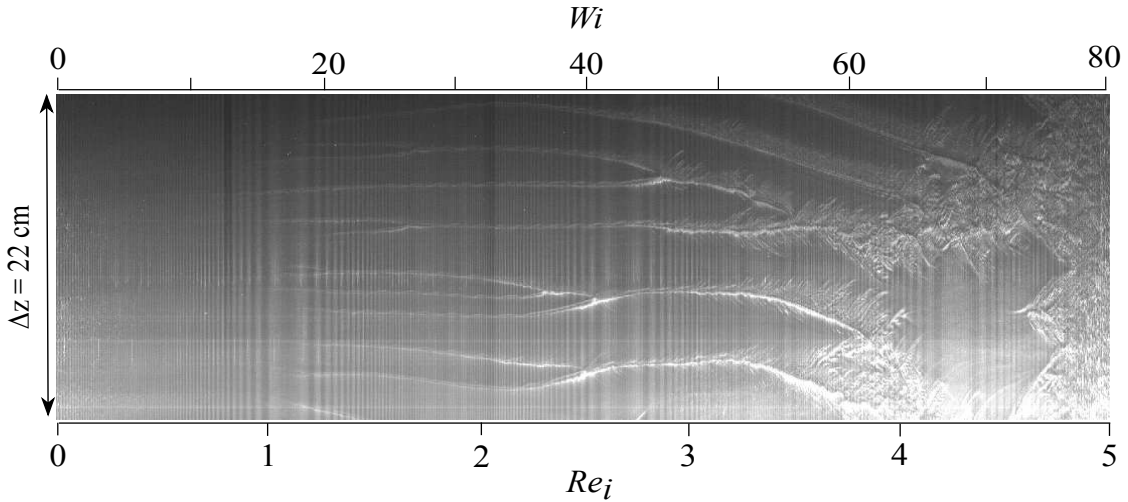}
   \caption{\small Space-time diagram of CTAB/NaSal at T = 22$^{\circ}$C {{\red during a gradual acceleration
   (0.001 rad/s$^{2}$) from rest to a shear rate of 17.7 s$^{-1}$. The vertical stripes in the image are caused by the temporal variation
   of the light source.}}}
   \label{fig2}
  \end{figure}
\subsubsection{Shear banding regime}

At low temperatures (e.g. T = 22$^{\circ}$C, T = 24$^{\circ}$C),
rheological measurements indicate that the wormlike micellar
solution exhibits shear banding. Fig.~(\ref{fig2}) shows the
space-time plot of this shear banding wormlike micellar fluid at T
= 22$^{\circ}$C with increasing $Re_{i}$. We report the initial
formation of stationary vortices at $Re_{i} = {{\red 0.85}}$, $Wi
= {{\red 15.8}}$, and $El = $ {{\red 18.6}} signaled by the
presence of horizontal lines on the space-time plot, and
characterized by an initial dimensionless axial wavelength of
$\approx$ 2.3. The dimensionless wavelength, $\lambda$, is defined
as the wavelength over the gap size $d$ and the wavelength is
calculated as the spacing between the bright bands shown in
Fig.~(\ref{fig2}). These bands merge and form stationary
structures characterized by a larger axial wavelength as we
increase the $Re_{i}$. Finally, we report a transition from evenly
spaced bright bands denoting stationary vortices to a chaotic
regime which is reminiscent of elastic turbulence at $Re_{i} =$
{{\red 2.9}}, $Wi = $ {{\red 44.9}} and $El =$ {{\red 15.5}}.
{{\red The vertical stripes evident in Fig.~(\ref{fig2}) and subsequent space-time plots are due to small
temporal fluctuations in the intensity of the light source.}}

\begin{figure}[h!]
  \centering
    \includegraphics [width=0.6\textwidth]{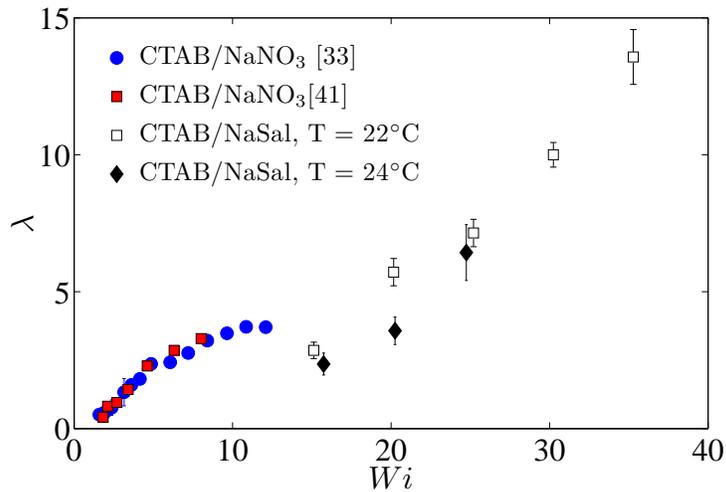}
   \caption{\small Dimensionless wavelength versus Weissenberg
    number for different wormlike  micelle solutions. }
   \label{fig3}
  \end{figure}

Fig.~(\ref{fig3}) shows the dimensionless wavelength ($\lambda$)
versus Weissenberg number ($Wi$) for our experiments on the
CTAB/NaSal system along with the results for the CTAB/NaNO$_{3}$
wormlike micellar fluid studied by Lerouge and co-workers. We also
include data from our own laboratory for the CTAB/NaNO$_{3}$
system~[44], which coincides with the published results  of
Lerouge and co-workers~\cite{Le08}. {{\red As shown in
Fig.~(\ref{fig3}), the results for CTAB/NaSal show some
differences with the CTAB/NaNO$_{3}$ system in terms of the
critical Wi for the onset of instability (inferred from
Fig.~(\ref{fig3})) and the wavelengths for the range of Wi
tested.}}\newline

The elasticity number for the transition from Couette flow to
stationary vortices as well as from stationary vortices to elastic
turbulence is high ($\gg1$) for the CTAB/NaSal system, and it
appears to follow the purely elastic pathway. {{\red The first transition occurs at an imposed
shear rate just above the start of the shear banding plateau.}} The purely elastic
criterion indicated in eq.~(\ref{eq1}) is modified by Fardin et
al. for shear-banding systems (cf. eq.~(1) in~\cite{Fa11}) to
account for the Weissenberg number $Wi_{h}$ in the high shear band
and the proportion of that band in the gap $\alpha$ as
$K_{c}\sim(\alpha\varepsilon)^{1/2}Wi_{h}$. Here, $Wi_{h}$ is
given by the longest relaxation time multiplied by the shear rate
in the high shear rate band. The proportion of the high shear band
is estimated using the lever rule: \beq \dot{\gamma} =
\alpha\dot{\gamma}_{h}+ (1-\alpha)\dot{\gamma}_{l},
  \label{eq5}
\eeq

where $\dot{\gamma}_{l}$ and $\dot{\gamma}_{h}$ are the shear
rates for onset of the shear banding plateau and the end of the shear
banding respectively. {{\red We note that this equation neglects the effects of shear-thinning.}}\newline
Our critical condition for the first
transition of  $Wi\approx $ {{\red 15.8}} at T = 22 $^{\circ}$C
corresponds to a critical condition
$K_{c}\sim(\alpha\varepsilon)^{1/2} Wi_{h} \approx {{\red 3}}$. Fardin et al.~\cite{Fa12E} showed that for the shear
banding system of CTAB/NaNO$_{3}$, the critical condition for
onset of elastic instability is $K_{c} \approx $ O(1). This is
also consistent with our measurements for the CTAB/NaNO$_{3}$
system, as suggested by Fig.~\ref{fig3}.\newline

Thus, the critical condition for onset of the elastic instability
in CTAB/NaSal differs {{\red modestly}} from that for the
CTAB/NaNO$_{3}$ system, and we note that at onset, the CTAB/NaSal
system displays a higher dimensionless wavelength than the
CTAB/NaNO$_{3}$ system. We note a few potential sources for these
discrepancies in $K_{c}$ and the wavelength at onset. First, wall
slip at the moving boundary might in principle affect the values
of $K_{c}$. It has been shown that the shear banding system based
on CTAB/NaNO$_{3}$ experiences significant wall slip at the onset
of shear banding~\cite{Fa12a}. Fardin et al.~\cite{Fa12E} showed
that the local $Wi_{h}$ measured via velocimetry techniques is
typically lower than the global $Wi_{h}$ measured with a
commercial rheometer for the beginning of the shear banding
regime. In particular, the local $Wi_{h}$ at the onset of shear
banding in their case is almost $1/3$ of the globally measured
$Wi_{h}$, and they used this locally measured $Wi_{h}$ to
calculate the critical threshold for onset of elastic instability.
Although there are no direct measurements of slip for the
CTAB/NaSal system, slip might potentially lower the effective
shear rate at the inner cylinder of our Taylor-Couette geometry
and this in turn would change the local $Wi_{h}$, the value of
$\alpha$ and thus $K_{c}$. Second, although the shear banding
wormlike micellar solution studied in this work at T =
22$^{\circ}$C shows no significant qualitative differences in
terms of non-linear rheology with the CTAB/NaNO$_{3}$ shear
banding wormlike system studied before~\cite{Fa12a,Fa12b},
quantitative differences exist. For example, the linear
viscoelasticity results for CTAB/NaSal are best fit with a two
mode Maxwell model, whereas, for CTAB/NaNO$_{3}$, a single mode
Maxwell model gives the best fit. The differences between the two
shear banding systems in linear viscoelasticity tests presumably
arise from different wormlike micellar structures. Therefore, in
addition to wall slip, this factor might also lead to differences
in the critical values $K_{c}$ and $\alpha$ at the onset of the
purely elastic instability for the two systems mentioned above.
{{\red We also note that, for polymer solutions, the value of the
critical condition and the axial wavelength of the purely elastic
instability are sensitive to viscosity ratio and to shear
thinning, with the dimensionless wavelength increasing with
increased shear thinning~\cite{La94}; thus, differences between
these quantities for the two wormlike micelle systems may also be
playing a role. }}
\subsubsection{Viscoelastic regime}
At higher temperatures, the CTAB/NaSal surfactant solution no
longer shows any sign of shear banding. Instead, the rheological
measurements show that it remains viscoelastic for the range of
temperature 26$^{\circ}$C $<$ T $<$ 32$^{\circ}$C. In this
intermediate elasticity regime, flow visualization in the
Taylor-Couette cell reveals new transitions that differ from those
of shear banding fluids and from those reported for weakly elastic
polymer solutions~\cite{Cr02,Du11}. Fig.~(\ref{fig4}) illustrates
regimes of the surfactant solution at different temperatures
within the range 26 - 32$^{\circ}$C, where the fluid shear-thins
but displays no shear-banding. {{\red As}} the temperature {{\red
is varied}} over this {{\red interval}} range, the elasticity
number {{\red changes}} from approximately 1 to 0.01.\newline

At T = 26$^{\circ}$C, {{\red where $El\approx 1$}}, travelling
waves that typically originate at the middle of the cylinders and
move towards both the top and bottom of the cylinders {{\red
form}} (c.f. Fig.~\ref{fig4}(a)) at $Re_{i} \approx {{\red 11.4}}$
and $Wi \approx {{\red 11.3}}$. The primary transition from
Couette flow is thus no longer to stationary secondary flows; we
refer to these structures as travelling vortices (TV). The
corresponding elasticity number for the onset of transition from
Couette flow to travelling vortices is about $El \approx {{\red
0.97}}$. To the best of our knowledge, this flow transition to
travelling vortices is reported for a surfactant based
viscoelastic fluid for the first time. {{\red With}} increasing
$Re_{i}$, these travelling vortices are replaced near $Re_{i}
\approx {{\red 27.2}}$, $Wi \approx {{\red 18.5}}$ by a chaotic
regime that is reminiscent of elastic turbulence. Perge et
al.~\cite{Pe14} also studied the transitions in flow of a
non-shear banding viscoelastic micellar system based on
CTAB/NaNO$_{3}$ in the same range of elasticitiy ($El \approx 1$)
and reported transitions as: CF $\rightarrow$ {{\red standing
waves}} (SW) $\rightarrow$ {{\red rotating standing waves}} (RSW)
$\rightarrow$ {{\red elastic turbulence}} (ET) with increasing
$Re_{i}$. These authors reported the critical condition for the
primary transition as $Re_{i} \approx 68$, $Wi \approx 75$ for
their geometry in which $\varepsilon = 0.087$. We can account for
the slight difference in curvature between their geometry and ours
(for which $\varepsilon = 0.0963$) by casting the critical
conditions in terms of inertial and elastic Taylor numbers where
$Ta_{i} = \varepsilon Re^{2}$ and $Ta_{e} = \varepsilon Wi^{2}$
respectively. However, even accounting for the small difference in
curvature, the critical conditions for the first transition (and
the form of the secondary flow) differ dramatically for the two
systems. We note that the viscosity ratio for the onset of
transition from CF$\rightarrow$ SW, in the system studied by Perge
et al.~\cite{Pe14} is approximately 50, whereas, in this study the
viscosity ratio for the onset of {{\red the}} first transition is
approximately {{\red 380}}. The results of Perge and co-workers
are similar to transitions reported for polymeric solutions based
on PEO/glycerol/water by Dutcher and Muller in the range of
elasticity of $El \approx 0.1- 0.2$ and viscosity ratio of
2.82~\cite{Du13}. However, a different transition sequence {{\red
is observed for the present CTAB/NaSal system}} at elasticity in
this range. This might be due to the difference between the
viscosity ratios{{\red, ramp protocol}} and/or the structure of
wormlike micelles studied in this paper and those studied by Perge
et al.~\cite{Pe14}. {{\red Perge et al.~\cite{Pe14} performed
their experiments by step increases in shear rate, whereas in this
study the ramp protocol is continuous and very slow (~0.001
rad/s$^{2}$). One would expect that a step or rapid ramp could
easily result in anomalously high critical conditions (if the flow
has insufficient time to develop) or forcing of the system into
alternate, nonlinear flow states.}}\newline

As the temperature {{\red is increased}} to T = 28$^{\circ}$C, the
first transition from Couette flow to disordered oscillations (DO)
is observed at $Re_{i} \approx {{\red 66.7}}$, for elasticity of
about $El \approx 0.2$ and the viscosity ratio of {{\red 130}}
(c.f. Fig.~\ref{fig4}(b) and Table~\ref{table}). For solutions of
linear, flexible polymers, the transition sequence in this range
of $El$ appears sensitive to the viscosity ratio
$\eta_{p}/\eta_{s}$. Groisman and Steinberg reported transition
from Couette flow to disordered oscillations for a shear thinning
viscoelastic solution based on PAAm/saccharose/water at $El
\approx 0.2-27$ and viscosity ratio of 0.82 ~\cite{St98}. However,
Crumeyrolle at al. reported a transition from Couette flow to
rotating standing waves in the moderate range of elasticity
(0.07-0.5) for a series of PEO/water solutions with viscosity
ratios in the range of 5 - 13~\cite{Cr02}. Additionally, Dutcher
and Muller, for $El \approx 0.1-0.2$, reported transitions from
Couette flow to stationary vortices to disordered rotating
standing waves to elastic turbulence for a PEO/glycerol/water
solution with a viscosity ratio of 2.8~\cite{Du13}. \newline
\begin{figure}[h!]
  \centering
    \includegraphics [width=0.93\textwidth]{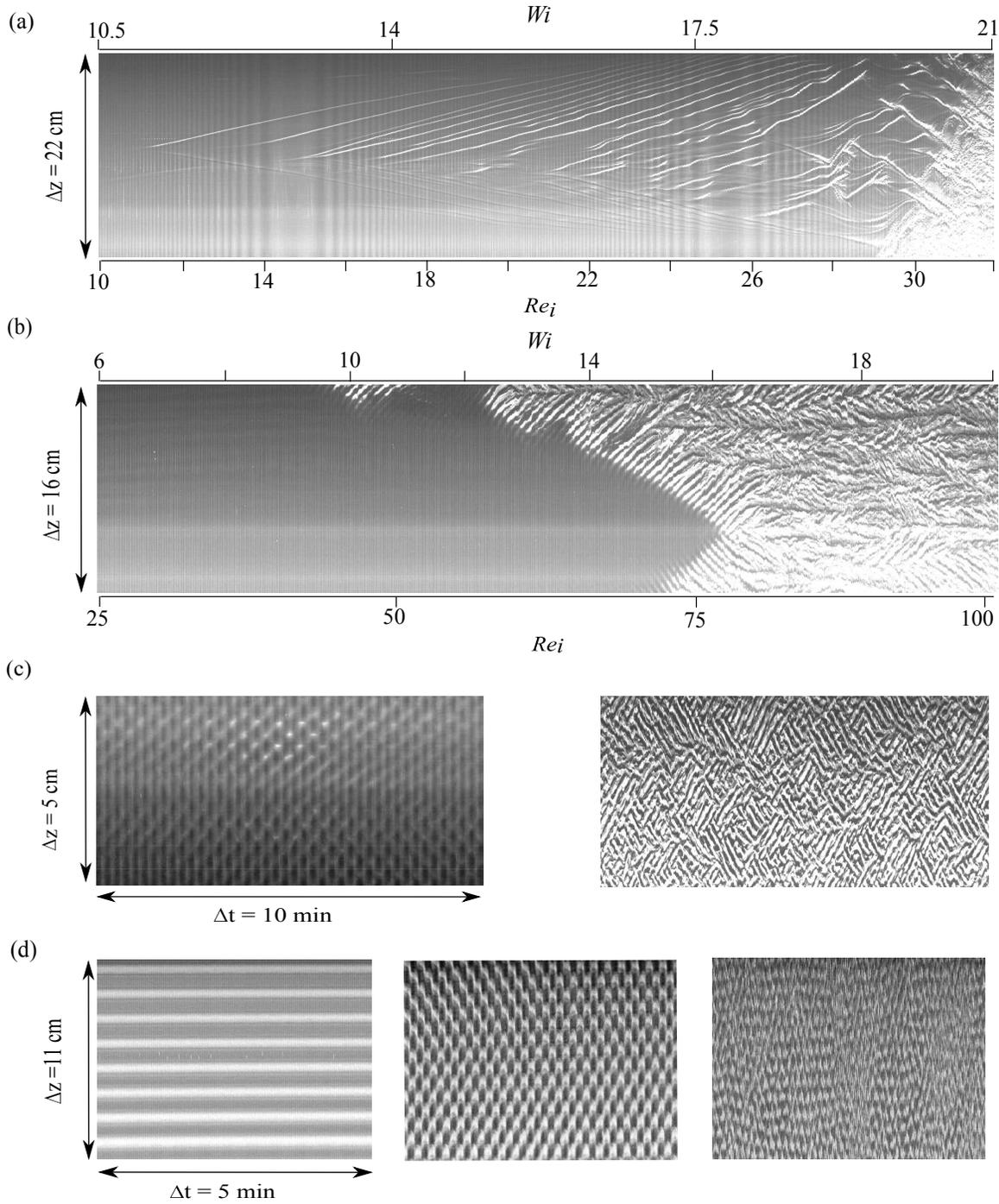}
   \caption{\small Space-time plots {{\red of CTAB/NaSal system at different temperature}} showing: (a) Couette flow to traveling vortices (${{\red Re_{i} \approx  11.4}}$ ) to elastic turbulence transition (${{\red Re_{i} \approx  27.2}}$ ) at
   T = 26$^{\circ}$C and $El \approx {{\red 0.97}}\rightarrow{{\red 0.72}}$.
   (b) Couette flow to disordered oscillations (${{\red Re_{i} \approx  66.7}}$) at T = 28 $^{\circ}$C and $El \approx 0.2$. (c) rotating standing waves ($Re_{i} \approx {{\red 79.7}}$ ) to disordered oscillations ($Re_{i} \approx {{\red 96.5}}$) at T = 30$^{\circ}$C and $El \approx {{\red 0.034}}$. (d) Taylor vortex flow ($Re_{i} \approx {{\red 142.7}}$) to rotating standing waves ($Re_{i} \approx {{\red 151.3}}$) to disordered oscillations ($Re_{i} \approx {{\red 179.5}}$) at T = 32$^{\circ}$C and $El \approx 0.01$.
   {{\red Experiments were conducted using a gradual acceleration (0.001 rad/s$^{2}$) from rest.}}}
   \label{fig4}
  \end{figure}

Fig.~\ref{fig4}(c) shows the rotating standing waves (RSW)
transition and the disordered oscillations (DO) at T =
30$^{\circ}$C. These transitions correspond to elasticity of about
$El \approx {{\red 0.034}}$ and the viscosity ratio {{\red 69}}.
Groisman and Steinberg showed for the system of
PAAm/saccharose/water in the range of 0.023 $ < El < $ 0.33 and
viscosity ratio of 0.78, that Couette flow is followed by rotating
standing waves and then disordered oscillations. Our results
indicate similar transitions for viscoelastic surfactant solutions
in this range of elasticity. However, our results seem to be at
odds with other experiments {{\red on linear, flexible polymer
solutions of PEO}} in the low elasticity range, which report a primary
transition to TVF, followed by wavy vortex flow~\cite{Cr02,Du11}.
\newline

Finally, at T = 32$^{\circ}$C, the surfactant solution is slightly
shear thinning (n = 0.95) {{\red and the following transition
sequences are observed}}: Couette flow to Taylor vortex flow to
rotating standing waves to disordered oscillations. This
experiment corresponds to elasticity of about $El \approx 0.01$
and the viscosity ratio of 39. For solutions of linear, flexible
polymers, the viscosity ratio again appears to play a role in the
transition sequence. A sequence similar to the one we observe for
our surfactant system was reported by Groisman and Steinberg for
$0.03 < El < 0.08$ and viscosity ratio of $\eta_{p}/\eta_{s}\sim
0.82$ for PAAM/saccharose-water solutions~\cite{Gr98a}. However,
Crumeyrolle et al. recovered {{\red the Newtonian, inertial
sequence of}} transitions for the range of $0.002 < El < 0.03$ and
viscosity ratios of $\eta_{p}/\eta_{s} = 0-3.45$ for PEO/water
solutions~\cite{Cr02}. In addition, Dutcher and Muller reported
primarily Newtonian transitions for $El\leq 0.023$ (Couette Flow
to TVF to WVF) with thresholds being slightly shifted due to
presence of elasticity~\cite{Du11}. {{\red Thus, while at
$El\sim 0.2$ (T = 28 $^{\circ}$C) the CTAB/NaSal solution displays a transition
sequence similar to that observed in PEO~\cite{Du13}, at 0.001 $<$ El $<$ 0.2 (28 $^{\circ}$C $<$ T $<$ 32 $^{\circ}$C),
the transition sequence follows that observed in PAAm solutions~\cite{Gr98a,St98}.
In both PEO and PAAm the sequence is sensitive to viscosity ratio, and the CTAB/NaSal viscosity ratio
is much larger than in earlier experiments with PEO and PAAm.}}
\newline

\subsubsection{Newtonian regime}

The surfactant solution exhibits nearly Newtonian behavior in
steady shear rheology at T = 35$^{\circ}$C (almost no discernible
shear-thinning and a very low relaxation time) and {{\red the
inertial sequence of}} transitions {{\red observed for Newtonian
fluids}} are also recovered in the flow visualization experiments
in the Taylor-Couette cell. These {{\red experiments}} correspond
to elasticity of about $El \approx 0.003$. Fig.~\ref{fig5} shows
space-time plots for the first three flow states following Couette
flow observed for the surfactant solution at T = 35$^{\circ}$C.
This is consistent with the results reported on both polymer
solutions and Newtonian fluids~\cite{Du11,Cr02}.
\begin{figure}[h!]
  \centering
    \includegraphics [width=1\textwidth]{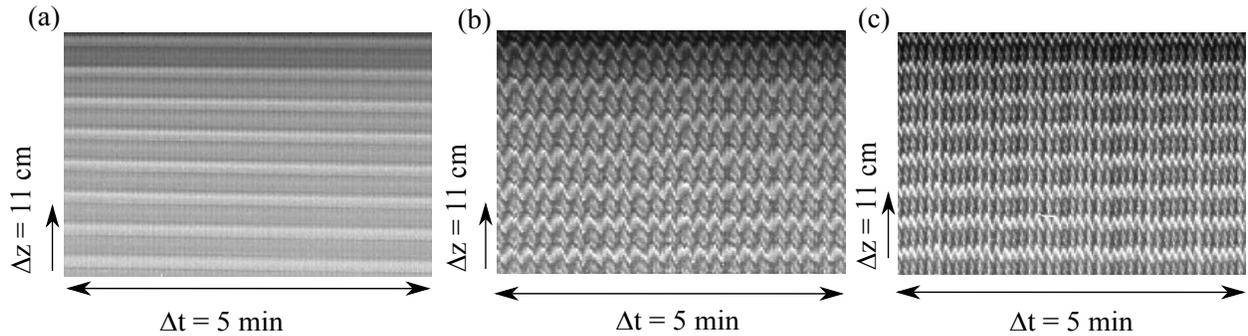}
   \caption{\small (a) Taylor Vortex Flow ($Re_{i} \approx 136$), (b) Wavy Vortex Flow ($Re_{i} \approx 157.3$) and (c) Modulated Wavy Vortex Flow ($Re_{i} \approx 181.5$) for the surfactant solution at T = 35$^{\circ}$C and $El \approx 0.003$.
   {{\red Experiments were conducted using a gradual acceleration (0.001 rad/s$^{2}$) from rest.}}}
   \label{fig5}
  \end{figure}\newline

Finally, Fig.~(\ref{fig6}) summarizes all of the above results in
terms of three dimensionless parameters: $Re_{i}/Re_{N}$, $Wi$ and
$El$. $Re_{N}$ denotes the onset of the TVF transition for a
Newtonian fluid in our Taylor-Couette cell ($Re_{N} \approx 140$).
In Fig.~(\ref{fig6}) we have also specified three regimes (I-III)
for which the surfactant solution shows different rheological
responses. In general, the rise in temperature from T =
22$^{\circ}$C to T = 35$^{\circ}$C lowers the elasticity (c.f.
Fig.~\ref{fig6}(a)) which in turn shifts the onset of
instabilities to higher $Re_{i}/Re_{N}$. As we approach the
Newtonian behavior (T$\rightarrow$35$^{\circ}$C), the onset of the
first transition gradually increases till it reaches values near
unity ($Re_{i}/Re_{N} \approx 0.97$). We also note that at the
lowest temperature - where the fluid is strongly shear thinning-
the critical elasticity numbers for the primary and secondary
transitions are varying as we ramp in $Re_{i}$ and as we increase
the temperature, this difference is weakened until it becomes
negligible at T = 32$^{\circ}$C {{\red where the shear-thinning is
very modest}}. However, increasing
the temperature has the opposite effect on the critical
Weissenberg number for the onset of instabilities. As the
temperature decreases (T$\rightarrow$22$^{\circ}$C), the critical
$Wi$ for onset of first transition increases and finally, in the
range of high elasticity (i.e. lowest temperature) when the
elastic instability is dominant, it reaches a constant value of
$Wi \approx {{\red 15-17}}$. This monotonic trend is consistent with prior
experimental results on PAAm/saccharose/water solutions by Groisman
and Steinberg for which an asymptotic value of $Wi \approx 25$ was
recovered~\cite{St98}.

\begin{figure}[h!]
  \centering
    \includegraphics [width=0.6\textwidth]{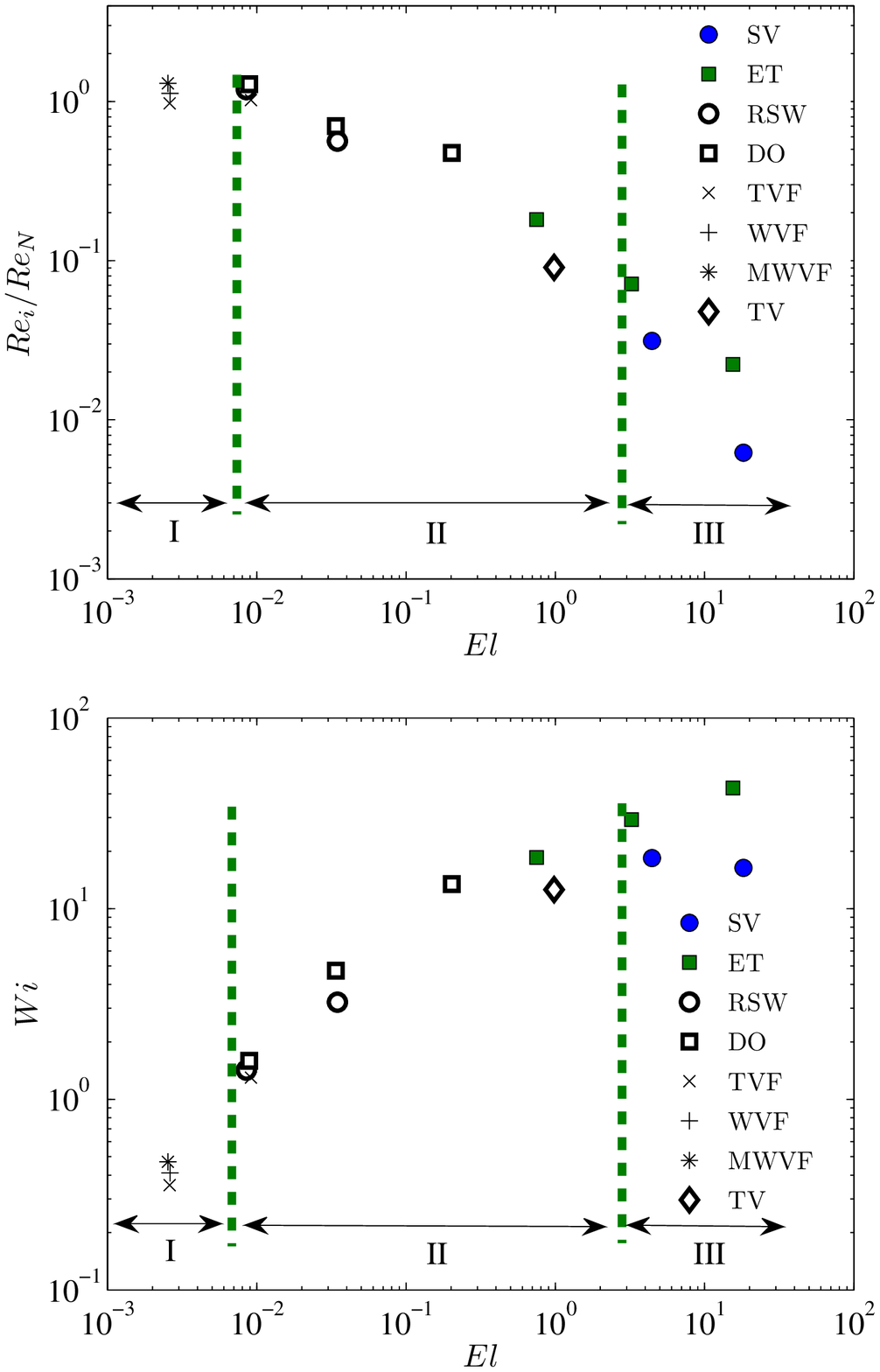}
   \caption{\small (a) Dimensionless Reynolds versus Elasticity number for different flow transitions with increasing $Re_{i}$.
   (b)  Weissenberg number versus Elasticity for all flow transitions. {{\red The following transitions: ${{\blu\bullet}}$ (stationary vortices)
   , ${{\blacksquare}}$ (elastic turbulence), $\circ$ (rotating standing waves), $\square$ (disordered oscillation), $\times$ (Taylor vortex flow), + (wavy vortex flow), $\ast$ (modulated wavy vortex flow) and $\lozenge$ (traveling vortices)
   are reported in the range of temperature 22-35 $^\circ$C.}} Regimes I-III correspond to Newtonian, shear thinning viscoelastic and shear banding rheology respectively. }
   \label{fig6}
  \end{figure}

\subsection{Effect of the Flow History}

Flow transitions and the final flow state at a particular set of
parameters for viscoelastic polymer solutions might in principle
be affected by the flow history. Such flow state hysteresis is
well-known for wavy vortex flow of Newtonian fluids as first
revealed by Coles~\cite{Co65}. Although hysteresis has been
reported in the Taylor-Couette flow of polymeric
solutions~\cite{St98,Du13}, wormlike micellar solutions based on
CTAB/NaNO$_{3}$ showed no sign of hysteresis in the studies of
Perge et al.~\cite{Pe14} for the moderate range of elasticity ($El
\approx 1$). We also report no sign of hysteresis for flow of
wormlike micelles based on CTAB/NaSal in the range of high and
moderate elasticity $El \geq 1$ for the transitions indicated in
Fig.~(\ref{fig6}). However, for elasticity of $El \approx 0.2$ and
$El \approx {{\red 0.034}}$ we report a transition similar to
solitary vortex pairs (diwhirls) as we decrease the $Re_{i}$.
Fig.~(\ref{fig7}) shows the transition from DO to oscillatory
strips to solitary vortex pairs (diwhirls) to Couette flow at T =
28$^{\circ}$C ($El\sim0.2$) as $Re_{i}$ decreases (compare this to
the pathways shown in Fig.~(\ref{fig6}) where $Re_{i}$ is
increasing). If the spacing between the bright bands (which appear
in Fig.~(\ref{fig7}) approximately over the range ${{\red 50}} <
Re_{i}< {{\red 28}}$) is small ($ < 5d$), they merge and form one
bright strip similar to the behavior of solitary vortex pairs
reported by Groisman and Steinberg~\cite{Gr97}. If the pairs are
far apart from each other ($ > 10d$), they do not interact with
each other. Here, $d$ represents the gap of our Taylor-Couette
cell ($d=0.7 cm$). For higher temperatures T $\geq $ 32$^{\circ}$C
that correspond to elasticity of $El < 0.01$, {{\red no}} signs of
hysteresis {{\red were observed}}.
\begin{figure}[h!]
  \centering
    \includegraphics [width=1\textwidth]{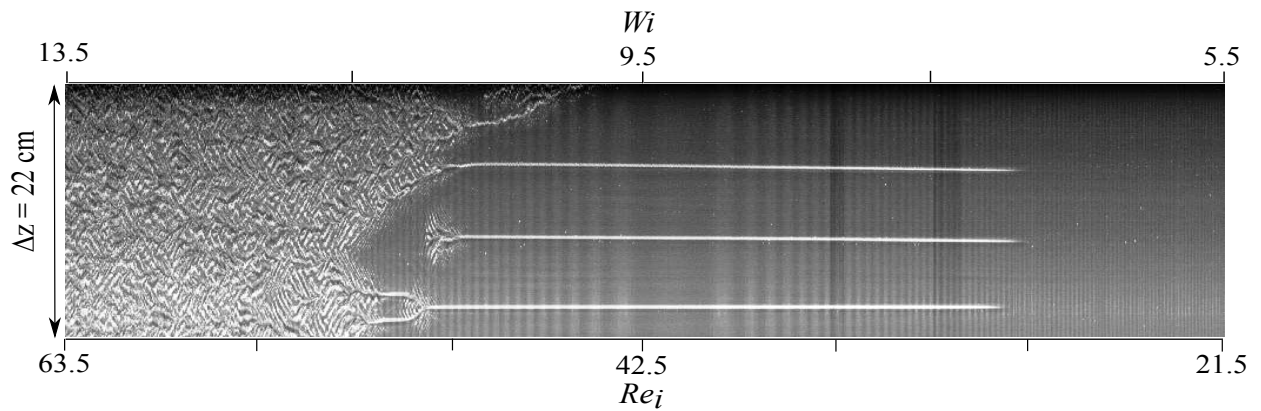}
   \caption{\small Space-time plot for the surfactant solution at T =28$^{\circ}$C, $El \approx 0.2$ with $Re_{i}$ {{\red(lower bar)}} and $Wi$ {{\red (upper bar)}} decreasing from left to the right.
   {{\red Experiments were conducted using a gradual deceleration (0.001 rad/s$^{2}$).}} }
   \label{fig7}
  \end{figure}

\subsection{Effects of Co-rotation and Counter-rotation}

Rotation of the outer cylinder can have a dramatic effect on flow
instabilities and transitions. Counter rotation of the cylinders
creates a nodal surface within the gap of the Taylor-Couette
geometry at some radial position $r = R_{n}$ (where $R_{i}< R_{n}<
R_{o}$) where the velocity is zero. This surface divides the gap
into an inner and an outer part. The inner part can be thought of
as a Taylor-Couette geometry defined by the rotating inner
cylinder and a stationary outer boundary defined by the nodal
surface. However, for the outer part, the inner boundary is
defined by the stationary nodal surface while the outer cylinder
is moving. Dutcher and Muller extensively studied co- and counter
rotation of the cylinders on flow transitions for a
PEO/glycerol/water solution at low elasticity $El = 0.023$. They
reported that both co- and counter rotation shifts the critical
Reynolds numbers $Re_{i}$ for the onset of both the primary and
secondary instabilities (to TVF and WVF, respectively) to higher
values, consistent with trends for Newtonian fluids~\cite{Du11}.
They also observed similar trends for the onset of the first two
instabilities in the range of moderate elasticity ($El \approx
0.1- 0.2$), although in this case the minimum in $Re_{i}$ is
shifted from $Re_{o} = 0$ to $Re_{o} \approx 30$ and the first
transition is to elastically modified stationary vortices, and the
second transition is to disordered rotating standing
waves~\cite{Du13}. For wormlike micelles however, there are no
experimental studies that examine the effect of rotation of the
outer cylinder. In the following experiments, the outer cylinder
{{\red is rotated}} quickly to reach its final value and then
{{\red rotation of}} the inner cylinder {{\red is gradually ramped
at a rate of 0.001 rad/$s^{2}$}}, increasing the inner cylinder
angular velocity until we reach the instability thresholds. For
co- and counter-rotation, we define $Re_{i} = \rho
R_{i}\Omega_{i}(R_{o} - R_{i}) /\eta(\dot{\gamma}_{i})$ and
$Re_{o} = \rho R_{o}\Omega_{o}(R_{o} - R_{i})
/\eta(\dot{\gamma}_{o})$ to be the Reynolds number at the inner
and outer cylinders respectively.
\begin{figure}[h!]
  \centering
    \includegraphics [width=1\textwidth]{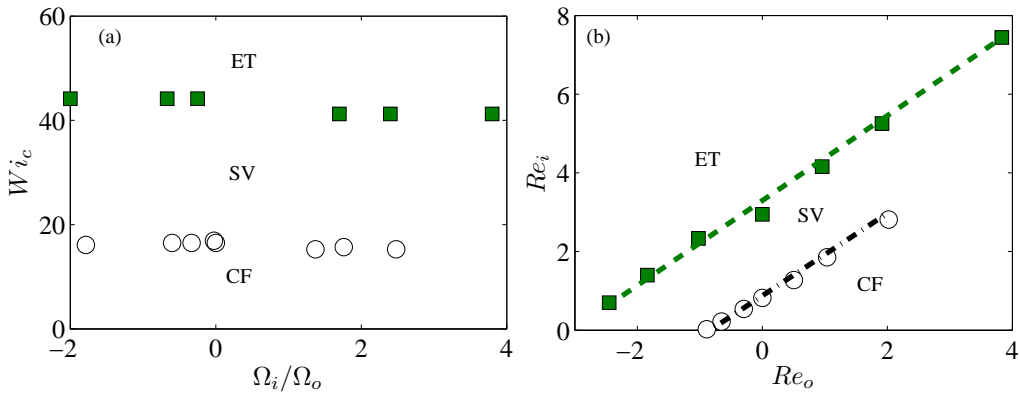}
   \caption{\small (a) Critical Weissenberg number versus rotation ratios for different transitions {{\red at T = 22$^{\circ}$C}}. (b) Critical Reynolds number at the inner cylinder versus Reynolds number at the outer cylinder for the CF ({{\red Couette flow}}) to SV ({{\red Stationary vortices}}) and ET ({{\red elastic turbulence}}) transitions {{\red at T = 22$^{\circ}$C}}.}
   \label{fig8}
  \end{figure}\newline

At high elasticity ($El \gg 1$, T = 22$^{\circ}$C), where the
surfactant solution exhibits shear banding, the instability is
purely elastic. Fardin et al.~\cite{Fa12a}, considering the case
of a stationary outer cylinder, showed that the critical threshold
for the onset of instability is set by the Weissenberg in the high
shear band for their system. We explore a wide range of rotation
ratios and report a constant Weissenberg number for onset of
instability, further supporting the idea that the instability is
purely elastic. Negative rotation ratios refer to counter rotation
whereas, positive ratios denote co-rotation of the cylinders.
Fig.~\ref{fig8}(a) shows the critical Weissenberg numbers for the
onset of stationary elastic vortices and elastic turbulence are
constant and about $Wi_{c} \approx {{\red 15.8\pm1}}$ and $Wi_{c}
\approx {{\red 42.8\pm2}}$, respectively. Fig.~\ref{fig8}(b)
indicates the critical Reynolds number associated with the two
transitions reported for shear banding wormlike micelles.
Fig.~\ref{fig8}(b) shows that the critical $Re_{i}$ increases with
increasing $Re_{o}$ such that the instability always starts at a
{{\red fixed}} critical shear rate (or Weissenberg number). We also calculated
the slopes of the critical curves in Fig.~\ref{fig8}(b) for the
two transitions. For transition from CF to SV the slope is
$\approx {{\red 0.97}}$ whereas for SV to ET transition the
magnitude of slope is $\approx {{\red 1.05}}$.\newline

At moderate elasticity ($El\approx1$), the viscoelastic surfactant
solution no longer exhibits shear banding and is still highly
shear thinning. In this range, we expect both inertia and
elasticity to play a substantial role in instabilities and
transitions. Fig.~\ref{fig9}(a) shows the critical Weissenberg
number for the onset of instabilities versus the rotation ratio.
For $\Omega_{i}/\Omega_{o} = 0$, ( i.e., a stationary inner
cylinder), the Weissenberg number for the first transition shows a
maximum ($Wi_{c} \approx {{\red 26.2}}$) and at high rotation
ratios (either positive or negative) the critical Weissenberg
number approaches an asymptotic value ($Wi_{c} \approx {{\red
12.2}}$, roughly consistent with the value reported above when
$\Omega_{o} = 0$, and $|\Omega_{i}/\Omega_{o}|\rightarrow\infty$).
Recalling that for a Newtonian fluid, the flow is stable when only
the outer cylinder is rotated {{\red until a catastrophic
transition to turbulence at $Re_{o}$ of order
40,000~\cite{Co65}}}, we expect that for $\Omega_{i}/\Omega_{o} =
0$, {{\red and the conditions considered here,}} inertial
destabilization plays no role and therefore, the flow will become
unstable via a purely elastic mechanism. Thus, the threshold of
instability should be set by a critical Weissenberg number. This
critical Weissenberg number ({{\red $Wi \approx 26.2$}}) is
somewhat higher than the value reported for onset of the purely
elastic instability at $El\gg 1$, possibly reflecting
stabilization of the flow by the higher solvent contribution to
the viscosity ratio. As the inner cylinder rotation increases from
zero, either via co- or counter-rotation, the effects of inertia
are apparent in the rapid decrease in the critical $Wi$ to $Wi
\approx {{\red 12.2}}$ for $|\Omega_{i}/\Omega_{o}|\approx 5$.
\begin{figure}[h!]
  \centering
    \includegraphics [width=1.0\textwidth]{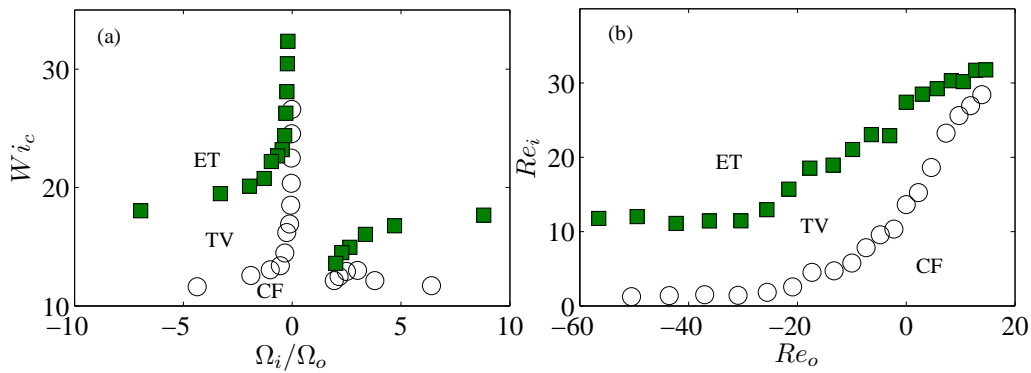}

   \caption{\small (a) Critical Weissenberg number versus rotation ratios for transition from CF ({{\red Couette flow}}) to TV ({{\red traveling vortices}}) at $El\approx {{\red 0.97}}$ and for TV to ET ({{\red elastic turbulence}}) at $El\approx0.72$ {{\red and T = 26$^{\circ}$C}}. (b) Reynolds number at the inner cylinder versus Reynolds at the outer cylinder for different transitions {{\red at T = 26$^{\circ}$C}}.}
   \label{fig9}
  \end{figure}
Fig.~\ref{fig9}(b) also shows the critical $Re_{i}$ versus
$Re_{o}$ for the two transitions for $El\approx1$. First, we note
that the critical $Re_{i}$ for the first transition, for all
$Re_{o}$ probed, is a small fraction of $Re_{i}$ for the Newtonian
case (where $Re_{i} \approx 140$ at $Re_{o} = 0$). In addition,
the first transition at $El \approx {{\red 0.97}}$ is to the
travelling wave structures identified above (cf.
Fig.~\ref{fig4}(a)) rather than to Taylor vortex flow (as reported
by Dutcher and Muller~\cite{Du11} for all $El \leq 0.023$) or
stationary vortices (reported by Dutcher and Muller~\cite{Du13}
for $El \approx 0.1-0.2$). The critical Reynolds {{\red number}} at the inner
cylinder for onset of travelling vortices decreases as we rotate
the cylinders in counter rotation fashion and finally approaches
an asymptotic value of $Re_{i} \approx {{\red 1.38}}$ for $Re_{o}
< {{\red -25}}$.\newline

At low elasticity ($El \approx {{\red 0.034}}$), inertial
destabilization becomes more dominant, but transitions are still
affected by the elasticity of the fluid. Therefore, we expect the
Reynolds number to be more relevant than a Weissenberg number in
this range of elasticity. Fig.~\ref{fig10}(a) shows that the
critical Weissenberg number for instability thresholds is smaller
than the ones reported in Fig.~\ref{fig9}(a), but the two figures
show similar trends with respect to the rotation ratio.
Counter-rotation of the outer cylinder reduces the Reynolds number
for the onset of instability and gives rise to a minimum value
near $Re_{o} \approx {{\red -54}}$. As $Re_{o}$ {{\red is decreased}} beyond
this value, the instability thresholds increase and at $Re_{o}
\approx -170$, the RSW state completely disappears and Couette
flow transitions directly to disordered oscillations. {{\red A similar}} trend
was also reported for polymeric solutions at low elasticity by
Dutcher and Muller ~\cite{Du11}. ~\nocite{Sup}
\begin{figure}[h!]
  \centering
    \includegraphics [width=1\textwidth]{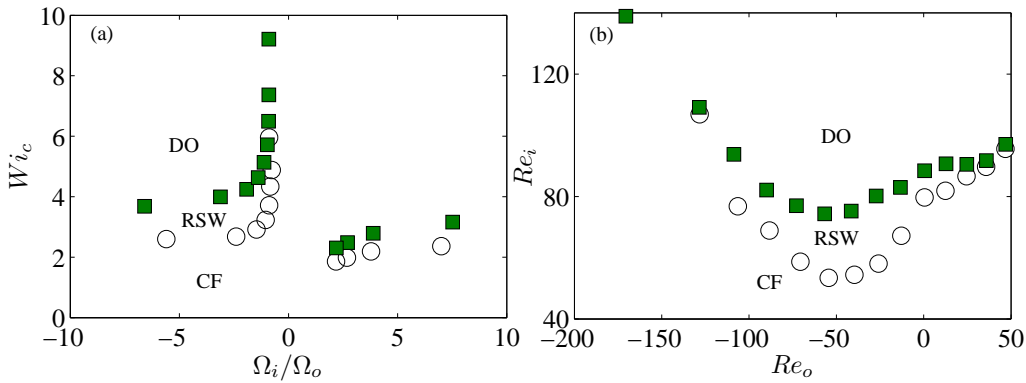}
   \caption{\small (a) Critical Weissenberg number versus rotation ratios for transitions {{\red CF (Couette flow), RSW (rotating standing waves) and DO (disordered oscillations)}} for $El\approx0.04$ {{\red at T = 30$^{\circ}$C}}. (b) Reynolds number at the inner cylinder versus Reynolds at the outer cylinder for different transitions {{\red at T = 30$^{\circ}$C}}.}
   \label{fig10}
  \end{figure}

\section{Conclusion}
In this work, we studied rheological and flow transitions in a
model wormlike micell{{\red e}} solution based on CTAB/NaSal over
a wide range of elasticity. The main experimental findings can be
summarized as below:\newline

At low temperatures (e.g. T = 22$^{\circ}$C and T =
24$^{\circ}$C), wormlike micelle solutions exhibit shear banding
in rheological measurements and in Taylor-Couette flow show the
formation of stationary elastic vortices beyond a critical
condition, $K_{C}\approx 3 $ followed by elastically dominated
turbulence as the flow strength is increased. The first transition
is characterized by an asymptotic wavelength which turns out to be
higher than the dimensionless wavelength reported for the shear
banding wormlike micelle system CTAB/NaNO$_{3}$ at the onset of
instability. One possibility is that the differences in critical
threshold for onset of instability in the shear banding fluid of
CTAB/NaSal ($K_{c} \approx 3$) and CTAB/NaNO$_{3}$ ($K_c \approx
1$) might potentially stem from wall slip at the inner cylinder.
Unfortunately, the turbidity contrast for this wormlike micellar
solution is not high enough to visualize the $r-z$ plane of the
Taylor-Couette cell and measurements of wall slip
 are beyond the scope of the present study. With regard to the difference in wavelength between the
 two wormlike micellar systems, we note that both experiments and theory for linear polymer solutions
 indicate that wavelength is sensitive to the details of the shear thinning~\cite{La94}. Moreover,
the shear banding wormlike micellar solution did not show any sign
of hysteresis. Rotation of both inner and outer cylinder yielded a
similar elastic instability that is characterized by a constant
Weissenberg number under these high elasticity conditions. \newline

In the intermediate range of temperature, where surfactant
solutions are viscoselastic and shear thinning, Taylor-Couette
experiments revealed a range of flow transitions and
instabilities. Starting with T = 26$^{\circ}$C, at a moderate
elasticity $El \approx 1$, we report a first transition from
Couette flow to the formation of vortices that are no longer
stationary and typically originate at the middle of the cylinders
and move towards both the bottom and the top of the cylinders.
To the best of our knowledge, this is a new transition for viscoelastic surfactant solutions.
These travelling waves are followed by a regime akin to elastic
turbulence. At this temperature, rotation of the cylinders in counter fashion lowers the critical
threshold for the onset of the travelling wave transition until
$Re_{i}$ reaches an asymptotic value of $Re_{i}\approx 1.38$ for
$Re_{o} < -25$. Increasing the temperature to T = 28$^{\circ}$C
reduces the viscosity and a transition from Couette flow to
disordered oscillation is reported at elasticity of $El \approx
0.2$. At T = 30$^{\circ}$C and $El \approx 0.034$, a transition
from Couette flow to rotating standing waves and disordered
oscillations is observed. Finally, at T = 32$^{\circ}$C, the
surfactant solution is slightly shear thinning and a transition
from Couette flow to TVF$\longrightarrow$RSW$\longrightarrow$DO is
observed. Most of the instabilities in this range of elasticity
are supercritical and show no hysteresis except for T =
28$^{\circ}$C and T = 30$^{\circ}$C for which disordered
oscillations are replaced by solitary vortex pairs as $Re_{i}$
decreases.\newline

At the highest temperature (T = 35$^{\circ}$C),
rheological measurements indicate nearly Newtonian behavior, and
flow visualization experiments in the Taylor-Couette cell also
reveal Newtonian transitions. Rotation of both cylinders shifts
the thresholds for onset of instability akin to results reported
for weakly elastic polymer solutions.\newline

Finally, we note that some of our results on this surfactant
solution are different from observations made on other wormlike
micelle solutions or polymer solutions. These differences may
arise from a number of factors including different range of
viscosity ratios, different extent of shear thinning, different
breakage times (in case of wormlike micelle solutions), different
extensional behaviors, differences in the wall slip, and
differences in the ramping protocols. In table~(\ref{table3}), we
summarize the transitions observed in the literature and in the
present work for both polymer solutions and wormlike micelle
solutions. As indicated in table~\ref{table3}, the range of
viscosity ratio reported in this paper (23 - 1340) is much higher
than those tested for polymer solutions (0 - 12.4). Other
rheological differences between solutions, e.g. differences in the
breakage time for wormlike micelle solutions, wall slip, and
differences in extensional behavior, are harder to measure or find
data in the literature. There are, at a minimum, qualitative
differences between the CTAB/NaSal system and the CTAB/NaNO$_{3}$
system even in the small amplitude oscillatory shear rheology.
This is evidenced by a different relaxation mechanism observed for
the surfactant solution based on CTAB/NaSal studied in this work
that follows a two-mode Maxwell model. The structural
characterization of this surfactant solution is beyond the scope
of the present work. Future work involves a through
characterization of the structure and extensional behavior of
these systems, and a better understanding of the origins of the
differences in flow behavior between wormlike micellar systems.

  \begin{longtable}{|c|c|c|c|c|c|c|c|}
        \hline
         Solution& $\eta_{p}/\eta_{s}$&$El$& \begin{tabular}{@{}c@{}@{}}Transition Sequence  \\ (increasing Re = $\uparrow$) \\ (decreasing Re = $\downarrow$) \end{tabular}& $Re_{o}$ & $R_{i}/R_{o}$ & $\frac{h}{(R_{o} - R_{i})}$ & Ref.\\
        \hline\hline
         \begin{tabular}{@{}c@{}@{}@{}@{}@{}@{}@{}@{}@{}} \\ \\  \\ PAAm/ \\ saccharose- \\water \\ \\ \\ \\ \\\end{tabular}   & \begin{tabular}{@{}c@{}@{}@{}@{}@{}@{}@{}@{}@{}} \\ 0.08 \\ \\ \hline~~0.78~~ \\\hline\\ 0.82 \\\\\\\end{tabular} & \begin{tabular}{@{}c@{}@{}@{}@{}@{}@{}@{}@{}@{}} 0.1-0.15\\ \hline 0.15-0.22\\ \hline 0.22-0.34\\\hline0.023-0.033\\\hline0.025\\\hline0.03-0.08\\\hline0.09-0.15\\\hline0.2-27\\  \end{tabular}  ~& \begin{tabular}{@{}c@{}@{}@{}@{}@{}@{}@{}} AZI$\uparrow$TVF$\uparrow$RSW$\uparrow$DO\\ \hline AZI$\uparrow$TVF$\uparrow$DO\\ \hline AZI$\uparrow$DO; \\DO$\downarrow$OS$\downarrow$DW$\downarrow$AZI\\\hline AZI$\uparrow$RSW$\uparrow$DO \\\hline AZI$\uparrow$TVF$\uparrow$WVF\\\hline AZI$\uparrow$TVF$\uparrow$RSW$\uparrow$DO\\\hline AZI$\uparrow$DO\\ \hline AZI$\uparrow$DO;\\DO$\downarrow$OS$\downarrow$DW$\downarrow$AZI \\ \end{tabular} ~& $0$ & \begin{tabular}{@{}c@{}@{}@{}@{}@{}@{}@{}@{}@{}} \\ 0.708 \\ \\\hline ~0.708~ \\\hline \\  \\0.829  \\ \\\end{tabular}   & \begin{tabular}{@{}c@{}@{}@{}@{}@{}@{}@{}@{}@{}} \\~~~ 54~~~ \\ \\\hline 54 \\\hline \\  \\74  \\ \\\end{tabular} & \begin{tabular}{@{}c@{}@{}@{}@{}@{}@{}@{}@{}@{}} \\ \cite{Gr96} \\ \\\hline ~~~\cite{Gr93} ~~~\\\hline \\  \\\cite{Gr98a,St98}  \\ \\\end{tabular} \\
          \hline
         \begin{tabular}{@{}c@{}} PEO/ \\ water \\\end{tabular}   & \begin{tabular}{@{}c@{}} 0~-~3.45\\\hline 5.32~-~12.4\\\end{tabular} & \begin{tabular}{@{}c@{}}0.1-0.15\\ \hline 0.15-0.22\\  \end{tabular}  ~& \begin{tabular}{@{}c@{}}~~~~ AZI$\uparrow$TVF$\uparrow$WVF~~~~\\ \hline ~~~~~~AZI$\uparrow$RSW~~~~~~\\ \end{tabular} ~& $0$ & $0.883$& $47$& $[12]$ \\
          \hline
         \begin{tabular}{@{}c@{}@{}@{}@{}@{}@{}@{}@{}@{}@{}c@{}@{}@{}@{}@{}@{}@{}@{}@{}} \\ \\ \\ \\ \\ \\ \\ PEO/ \\ glycerol-\\water \\ \\ \\ \\ \\ \\ \\ \\ \\ \\ \\ \end{tabular}   & \begin{tabular}{@{}c@{}@{}@{}@{}@{}@{}@{}@{}@{}@{}c@{}@{}@{}@{}@{}@{}@{}@{}@{}} \\ \\ 0.3\\ \hline  \\ \\ \\0.93 \\ \hline  \\ \\ \\ 0.92 \\ \hline  \\ \\0.78 \\ \hline \\ \\  2.82 \\ \\ \\ \\ \end{tabular} & \begin{tabular}{@{}c@{}@{}@{}@{}@{}@{}@{}@{}@{}@{}c@{}@{}@{}@{}@{}@{}@{}@{}@{}} \\ \\ 0.00047\\ \hline  \\ \\ \\0.0017 \\ \hline  \\ \\ \\ 0.0054 \\ \hline  \\ \\0.023 \\ \hline \\ \\ $\sim$0.1-0.2 \\ \\ \\ \\ \end{tabular} & \begin{tabular}{@{}c@{}@{}@{}@{}@{}@{}@{}@{}c@{}@{}@{}@{}@{}@{}@{}} AZI$\uparrow$TVF$\uparrow$WVF$\uparrow$MVF$\uparrow$TTV\\ \hline AZI$\uparrow$TVF$\uparrow$WVF\\ \hline AZI$\uparrow$TVF$\uparrow$WVF(h)\\ \hline AZI$\uparrow$TVF$\uparrow$WVF$\uparrow$MVF$\uparrow$WVF \\ MWV$\uparrow$CWV$\uparrow$WTV$\uparrow$MT \\\hline AZI$\uparrow$TVF$\uparrow$WVF\\\hline AZI$\uparrow$TVF$\uparrow$WVF(h)\\\hline AZI$\uparrow$TVF$\uparrow$WVF$\uparrow$MVF \\$\uparrow$WVF$\uparrow$WTV(El)/CWV(El) \\ \hline AZI$\uparrow$TVF$\uparrow$WVF\\\hline AZI$\uparrow$TVF$\uparrow$WVF(h)\\\hline AZI$\uparrow$TVF$\uparrow$WVF$\uparrow$MVF$\uparrow$WVF \\\hline AZI$\uparrow$TVF$\uparrow$WVF\\\hline AZI$\uparrow$TVF$\uparrow$WVF(h)\\ \hline AZI$\uparrow$SV$\uparrow$DRSW$\uparrow$EDT\\ EDT$\downarrow$DRSW$\downarrow$RSW$\downarrow$AZI\\ \hline AZI$\uparrow$SV$\uparrow$DRSW$\uparrow$EDT\\ EDT$\downarrow$DRSW$\downarrow$RSW$\downarrow$AZI \\\hline AZI$\uparrow$DRSW$\uparrow$EDT\\\hline AZI$\uparrow$EDT\\ \end{tabular} ~&  \begin{tabular}{@{}c@{}@{}@{}@{}@{}@{}@{}@{}@{}@{}@{}@{}@{}@{}@{}@{}@{}@{}@{}} 0\\\hline -100--100\\ \hline ~200--500~ \\ \hline \\ 0 \\\hline -100--100 \\ \hline 200  \\ \hline 0 \\ \\ \hline -100--0\\ \hline 200--400 \\ \hline 0 \\ \hline -100--100 \\ \hline 140--200\\ \hline 0\\ \\ \hline-100--150 \\ \\ \hline-180-- -100 \\ \hline -225\\   \end{tabular}   & \begin{tabular}{@{}c@{}@{}@{}@{}@{}@{}@{}@{}@{}} \\ \\ \\ \\ \\ \\ 0.912 \\ \\ \\ \\ \\ \\  \\  \\ \hline \\ \\ \\ 0.912\\ \\ \\ \end{tabular}& \begin{tabular}{@{}c@{}@{}@{}@{}@{}@{}@{}@{}@{}} \\ \\ \\ \\ \\ \\ 60.7 \\ \\ \\ \\ \\ \\  \\  \\ \hline \\ \\ \\ 60.7\\ \\ \\ \end{tabular}& \begin{tabular}{@{}c@{}@{}@{}@{}@{}@{}@{}@{}@{}} \\ \\ \\ \\ \\ \\ \cite{Du11} \\ \\ \\ \\ \\ \\  \\  \\ \hline \\ \\ \\ \cite{Du13}\\ \\ \\ \end{tabular} \\
          \hline
         \begin{tabular}{@{}c@{}}   CTAB/NaNO$_{3}$ \\ \end{tabular}   & \begin{tabular}{@{}c@{}}  40 - 50 \\ \end{tabular} & \begin{tabular}{@{}c@{}}  $\sim$0.8-1.1 \\
         \end{tabular} & \begin{tabular}{@{}c@{}} AZI$\uparrow$SV$\uparrow$DRSW$\uparrow$ET \\ \hline ET$\downarrow$DRSW$\downarrow$SV$\downarrow$AZI\\ \end{tabular}
         &  \begin{tabular}{@{}c@{}}  \\ 0 \\ \end{tabular} & \begin{tabular}{@{}c@{}}  \\ 0.92 \\ \end{tabular} & \begin{tabular}{@{}c@{}}  \\ 30 \\ \end{tabular}
         & \begin{tabular}{@{}c@{}}  \\ \cite{Pe14} \\ \end{tabular}
         \\

         \hline
            \begin{tabular}{@{}c@{}}  CTAB/NaNO$_{3}$ \\ \end{tabular} & \begin{tabular}{@{}c@{}}   $> 10^{3}$ \\ \end{tabular} & \begin{tabular}{@{}c@{}}   $> 10^{3}$ \\ \end{tabular}
         & \begin{tabular}{@{}c@{}}   AZI$\uparrow$ZZ$\uparrow$AF$\uparrow$SV$\uparrow$F$\uparrow$ET \\ \hline AZI$\uparrow$MAF$\uparrow$MF$\uparrow$ET\\ \end{tabular}
         & \begin{tabular}{@{}c@{}}  0 \\ \end{tabular} &  \begin{tabular}{@{}c@{}}  0.83-0.96 \\ \hline 0.53\\ \end{tabular} &  \begin{tabular}{@{}c@{}}  16-40 \\ \hline 3.42\\ \end{tabular} &
         \begin{tabular}{@{}c@{}}  \cite{Fa12b} \\  \end{tabular}
         \\
          \hline

              \begin{tabular}{@{}c@{}}  CPCl/NaSal- \\ NaCl \end{tabular} & \begin{tabular}{@{}c@{}}   $> 10^{3}$ \\ \end{tabular} & \begin{tabular}{@{}c@{}}   $ 6\times10^{3}$\\ $-2\times10^{4}$  \end{tabular}
           & \begin{tabular}{@{}c@{}}   AZI$\uparrow$SV$\uparrow$ET \\  \end{tabular}
           & \begin{tabular}{@{}c@{}}  0 \\ \end{tabular} &  \begin{tabular}{@{}c@{}}  0.92 \\  \end{tabular} &  \begin{tabular}{@{}c@{}}  35.4\\ \end{tabular} &
           \begin{tabular}{@{}c@{}}  \cite{Fa12a} \\  \end{tabular}
           \\
            \hline

    \begin{tabular}{@{}c@{}}  CTAB/NaNO$_{3}$ \\ \end{tabular} & \begin{tabular}{@{}c@{}}   $2.2\times10^{3}$ \\ $-1.9\times10^{4}$\\ \end{tabular} & \begin{tabular}{@{}c@{}}   8.5-73.5\\  \end{tabular}
 & \begin{tabular}{@{}c@{}}   AZI$\uparrow$SV$\uparrow$ET \\  \end{tabular}
 & \begin{tabular}{@{}c@{}}  0 \\ \end{tabular} &  \begin{tabular}{@{}c@{}}  0.912 \\  \end{tabular} &  \begin{tabular}{@{}c@{}}  60.7\\ \end{tabular} &
 \begin{tabular}{@{}c@{}}  \cite{Mo16a} \\  \end{tabular}
 \\
  \hline

 \begin{tabular}{@{}c@{}@{}@{}@{}@{}@{}@{}@{}@{}@{}} \\ \\  \\ \\  \\CTAB/NaSal \\ \\ \\ \\ \\ \\\end{tabular}   & \begin{tabular}{@{}c@{}@{}@{}@{}@{}@{}@{}@{}@{}} \\  400-1340 \\\hline  \\ 270-380 \\\hline  \\ 130 \\\hline \\ 70 \\ \\\hline 39 \\ \hline 23 \\ \end{tabular} & \begin{tabular}{@{}c@{}@{}@{}@{}@{}@{}@{}@{}@{}@{}} \\  3.3-18.6 \\\hline  \\ 0.72-0.99 \\\hline  \\ 0.2 \\\hline \\ 0.034 \\ \\\hline 0.01 \\ \hline 0.003 \\ \end{tabular}  ~& \begin{tabular}{@{}c@{}@{}@{}@{}@{}@{}@{}@{}} AZI$\uparrow$SV$\uparrow$ET\\ \hline AZI$\uparrow$SV$\uparrow$ET\\ \hline AZI$\uparrow$TV$\uparrow$ET \\\hline AZI$\uparrow$TV$\uparrow$ET\\\hline AZI$\uparrow$DO \\ DO$\downarrow$OS$\downarrow$DW$\downarrow$AZI\\\hline AZI$\uparrow$RSW$\uparrow$DO\\\hline AZI$\uparrow$RSW$\uparrow$DO\\ \hline AZI$\uparrow$DO\\\hline AZI$\uparrow$RSW$\uparrow$DO \\ \hline AZI$\uparrow$TVF$\uparrow$WVF$\uparrow$MWVF\\ \end{tabular} ~& \begin{tabular}{@{}c@{}@{}@{}@{}@{}@{}@{}@{}@{}} 0 \\\hline -1--2 \\ \hline 0 \\\hline -50--20 \\\hline  \\0 \\ \hline 0  \\ \hline -140--50  \\\hline $<$ -140 \\\hline 0\\\hline 0\\  \end{tabular} & $0.912$& $60.7$& \begin{tabular}{@{}c@{}@{}@{}@{}@{}@{}@{}@{}@{}@{}} \\ \\ \\ \\ \\  This \\work \\ \\ \\ \\ \\\end{tabular} \\
  \hline

    \caption{\small Comparison of transition sequences in Taylor-Couette flows of linear flexible polymer solutions and wormlike micelle solutions.
    The flow state ZZ, AF, F, MAF and MF denote zig-zag, anti-flame, flame, modified anti-flame and modified flame states for wormlike micelle solutions. }

    \label{table3}
  \end{longtable}

\section{Supplementary material}
See supplementary material for experiments that show more detailed
information on CTAB/NaSal and CTAB/NaNO$_{3}$ systems.

\section{Acknowledgement}
The authors gratefully acknowledge the financial support from
ACS-PRF 53143-ND9 and also NSF through grant number CBET 1335653.
The authors are also grateful to Malvern Instrument for the loan
of the Malvern Gemini rheometer.

\bibliographystyle{unsrt}


\end{document}